\documentclass[12pt]{article}
\usepackage{amsfonts}
\usepackage{color,caption}
\usepackage{graphicx}
\usepackage{amsmath}
\usepackage{float}
\usepackage{amssymb}
\providecommand{\U}[1]{\protect\rule{.1in}{.1in}}
\textwidth=6.5in \hoffset=-.75in \voffset=-1in \numberwithin{equation}{section}
\numberwithin{figure}{section}

\newcommand {\be}{\begin{equation}}
 \newcommand {\ee}{\end{equation}}
 \newcommand {\bea}{\begin{eqnarray}}
 
 \newcommand {\eea}{\end{eqnarray}}

\def \th {\theta}
\def \O {\Omega}
\def \f {\frac}
\def \mb {\mathbb}
\def \D {\Delta}
\def \sr {\sqrt}
\newcommand{\rn}{Reissner-Nordstrom~}
\newcommand{\sch}{Schwarzschild~}

\begin{document}

\begin{titlepage}
\bigskip \begin{flushright}
\end{flushright}
\vspace{1cm}
\begin{center}
{\Large \bf {Deformed Hidden Conformal Symmetry for Rotating Black Holes}}\\
\vskip 1cm
\end{center}
\vspace{1cm}
\begin{center}
A. M.
Ghezelbash{ \footnote{amg142@campus.usask.ca}}, H. M. Siahaan{\footnote{hms923@campus.usask.ca}}
\\
Department of Physics and Engineering Physics, \\ University of Saskatchewan,
Saskatoon, Saskatchewan S7N 5E2, Canada\\
\vspace{1cm}
\end{center}

\begin{abstract}

The generic non-extremal Kerr-Newman black holes are holographically dual to hidden conformal field theories in two different pictures. The two pictures can be merged together to the CFT duals in general picture that are 
generated by $SL(2,\mathbb{Z})$ modular group.
We find some extensions of the conformal symmetry generators that yield an extended local family of $SL(2,\mathbb{R})_L \times SL(2,\mathbb{R})_R$ hidden conformal symmetries for the Kerr-Newman black holes 
parameterized by one deformation parameter.  The family of deformed hidden conformal symmetry for Kerr-Newman black holes also provides a set of deformed hidden conformal generators for the charged Reissner-Nordstrom black holes. The set of deformed hidden conformal generators reduce to the hidden $SL(2,\mathbb{R})$ conformal generators for the Reissner-Nordstrom black hole for specific value of deformation parameter.
We also find agreement between the macroscopic and microscopic entropy and absorption cross section of scalars for the Kerr-Newman black hole by considering the appropriate temperatures and central charges for the deformed CFTs. 

\end{abstract}
\end{titlepage}\onecolumn
\bigskip

%\tableofcontents

\section{Introduction}
The Kerr/CFT correspondence relates the physics of the extremal Kerr black hole to that of a dual chiral conformal field theory (CFT) \cite{Compere,CargeseKerrCFT}. The generators of CFT related to the diffeomorphisms of the near horizon geometry of the Kerr black hole \cite{stro}. The correspondence has been established for different types of extremal rotating black holes in four dimensions as well as higher dimensional black holes \cite{KerrCFT1}-\cite{KerrCFT4}.

For the more interesting non-extremal black holes, one may consider a different approach to establish the correspondence \cite{Castro}. For these generic rotating black holes, the conformal symmetry that is know as the hidden conformal symmetry can be found by looking at the solutions of a test field in the background of the black hole. In general, the test field is considered to be a neutral scalar field. It turns out that one can construct proper generators for an $SL(2,\mathbb{R})$ algebra such that its Casimir invariant produces exactly the wave equation of the neutral scalar field. The hidden conformal symmetry for the wave equation of scalar field in  different rotating black holes was obtained explicitly in \cite{othe1,othe2,KSChen}.
Moreover, it is shown that for some charged rotating black holes, there are more than one dual CFT. As an example, for the four-dimensional Kerr-Newman black holes, there is one class of CFT that is associated to the rotation of black hole. The second class of CFT is associated to the charge of black hole. The former and latter CFTs are called CFTs in $J$ and $Q$ pictures, respectively \cite{Chen-2fold,chen-novelsl2z,Chen-genhidden4d5d}. 

We also notice that there is correspondence between the angular momentum of the Kerr-Newman black hole to the rotational symmetry of black hole in $\phi$ direction. Moreover the charge of the black hole is in correspondence to the gauge symmetry of the black hole. The gauge symmetry of the black hole can be  interpreted as the rotational symmetry of the uplifted black hole in the fifth-direction $\chi$. These facts lead to two new CFTs correspond to two rotational symmetries of the uplifted black hole in five dimensions and we call them  $\phi '$ and $\chi '$ pictures respectively \cite{Chen-genhidden4d5d}. Using the modular group $SL(2,\mathbb{Z})$ for a torus made by $(\phi,\chi)$, we get one other picture that we call as the general picture. We also note that the $J$ and $Q$ pictures for the Kerr-Newman black hole can be obtained by choosing the identity element of the modular group in the general picture \cite{Chen-genhidden4d5d}.

As the second example, we consider the class of rotating charged Kerr-Sen black holes in four dimensions \cite{sen}. The Kerr-Sen solutions are the exact solutions of the low energy limit of 
heterotic string theory in four dimensions. The solutions include three non-gravitational fields
that do not contribute to the central charges of the dual chiral CFT \cite{GH2}.
However, unlike the Kerr-Newman black hole, the Kerr-Sen black hole does not possess any well defined $Q$ picture. The only consistent picture for Kerr-Sen black hole is the $J$ picture \cite{Ghezelbash}.
The other feature of the Kerr-Sen black hole is that the hidden conformal symmetry generators can be extended by including a deformation parameter in the radial equation of motion. These extended (deformed) hidden conformal symmetries in the limit of zero rotation provide the hidden conformal symmetry for the charged Gibbons-Maeda-Garfinkle-Horowitz-Strominger black holes \cite{GMGHS}. The idea of introducing the deformation parameter in the radial equation of motion first discussed in \cite{Lowe} to extend the hidden conformal symmetry of the non-extremal Kerr black hole. One direct benefit of introducing the deformation parameter in the radial equation of motion of scalar field is that one can find the hidden conformal symmetry for the non-rotating black hole from the deformed hidden conformal symmetry of the rotating black hole in the limit of zero rotation.  As the first example, the hidden conformal symmetry of Schwarzchild black hole can be obtained from the deformed hidden conformal symmetry of Kerr black hole for special value of deformation and rotational parameters \cite{Lowe}. Moreover, the hidden conformal symmetry for the charged non-rotating Gibbons-Maeda-Garfinkle-Horowitz-Strominger black hole can be obtained directly from the deformed hidden conformal symmetry of Kerr-Sen black hole for  special value of deformation parameter and in the limit of zero rotation \cite{GMGHS}.  We also note that we choose the deformation parameter in such a way that the low frequency limit and the near region geometry remain intact.  The physical justification for the deformation in the radial equation near the inner horizon is related to the fact that the solutions to the exact radial equation for the scalar field are singular at the inner horizon, outer horizon and far infinity. However the back-reaction of the field on the internal geometry of black hole replaces the inner Cauchy horizon by a null curvature spacelike singularity that covers the inner horizon of the black hole \cite{back}. As a result, the region behind the null spacelike singularity that includes the inner horizon is not the physical region of interest in the solutions of the radial equation. The other interesting feature of deformation of the inner horizon is that it doesn't change the location of other singularities of the radial equation that are located on the outer horizon and far infinity.  

Inspired by comparing the four-dimensional charged rotating Kerr-Sen black holes to the Kerr-Newman black holes, the lack of $Q$ picture for the Kerr-Sen black holes and existence of the extended (deformed) hidden conformal symmetries for the Kerr-Sen black holes, in this article
we investigate the existence of the extended (deformed) hidden conformal symmetry for the Kerr-Newman black holes in different pictures. 

The article is organized as follows: In section \ref{gen-hidden}, we briefly review the hidden conformal symmetries for the Kerr-Newman black holes in different pictures. In section \ref{deformedJ}, we introduce the deformed radial function and find explicitly the deformed hidden conformal generators for the Kerr-Newman black holes in the $J$ picture. We also find that the deformed right and left temperatures and central charges of the dual CFT lead to the microscopic entropy of the CFT that is in exact agreement with the macroscopic entropy. Then in section \ref{deformedQ}, we construct the deformed hidden conformal generators for the Kerr-Newman black holes in the $Q$ picture. We again find that the deformed right and left temperatures and central charges of the dual CFT lead to the microscopic entropy of the CFT that is in exact agreement with the macroscopic entropy. In section \ref{deformedGen}, we find the deformed radial equation in general $\phi '$ picture and find the deformed hidden conformal generators for the Kerr-Newman black holes. We again find that the deformed right and left temperatures and central charges of the dual CFT lead to the microscopic entropy of the CFT that is in exact agreement with the macroscopic entropy in $\phi '$ picture. In section \ref{scat}, we consider the scattering of charged scalars in the background of Kerr-Newman black holes based on deformed radial equation in three different pictures. The results in all three pictures provide support for correspondence between the Kerr-Newman black holes and dual CFTs in different pictures. In section \ref{reis}, we find the deformed hidden conformal generators for the \rn black holes. Finally in section \ref{conc} we wrap up the article by concluding remarks.

\section{The hidden conformal symmetries for Kerr-Newman black holes}\label{gen-hidden}

The Kerr-Newman metric is the exact solution to Einstein-Maxwell equations. The solution describes a spacetime outside of an electrically charged rotating massive object. The line element of Kerr-Newman spacetime can be read as \cite{Chen-2fold,Chen-genhidden4d5d} 
\be
ds^2 = - \frac{\Delta - a^2 \sin^2\theta}{\varrho} \left[ dt \!+\! \frac{(2 M r \!-\! Q^2) a \sin^2\theta}{\Delta - a^2 \sin^2\theta} d\phi \right]^2 + \varrho  \frac{dr^2}{\Delta} + \varrho d\theta^2 + \frac{\varrho\Delta \sin^2\theta}{\Delta - a^2\sin^2\theta} d\phi^2,\label{KNmetric}
\ee
where
\be
\varrho=r^2+a^2\cos^2\th ,\ee
\be
\D=r^2-2Mr+a^2+Q^2.\label{deltainkerr}
\ee
The metric (\ref{KNmetric}) in the limit of $a=0$ reduces to the Reissner-Nordstrom solution, whereas in the limits $a=0,~Q=0$, the metric (\ref{KNmetric}) reduces to Schwarzschild spacetime. The gauge fields in the solution
are
\be A = - \frac{Q r}{\varrho} \left( dt - a \sin^2\theta d\phi \right). \ee
The inner and outer horizons, $r_-$ and $r_+$ respectively, can be expressed as
\be
r_{\pm}=M\pm\sr{M^2-a^2-Q^2}.
\ee
\par
For the extremal Kerr-Newman black holes, $M^2 = a^2 + Q^2$ which provides $r_+ = r_- = M$. The Bekenstein-Hawking entropy, Hawking temperature, angular velocity and the electric potential at the horizon of the black hole (\ref{KNmetric}) can be read as
\begin{eqnarray} \label{en}
&&S_{BH}=\pi (r_+^2+a^2),\label{en1}\\
&&T_H=\f{r_+-r_-}{4\pi(r_+^2+a^2)},\label{en2}\\
&&\O_H=\f{a}{r_+^2+a^2},\label{en3}\\
&&\Phi_H=\f{Qr_+}{r_+^2+a^2},\label{en4}
\end{eqnarray}
respectively. \par
We consider a massless charged test scalar field in the background of the Kerr-Newman black
hole.  The minimally coupled equation of motion for the scalar field is 
\begin{equation}
(\nabla_{\alpha} - i e A_{\alpha})(\nabla^{\alpha} - i e A^{\alpha})
\Phi = 0\label{KG1},
\end{equation}
where $e$ is the electric charge of scalar field. There are two  Killing vectors $\partial_t$ and $\partial_\phi$ for the Kerr-Newman black holes (\ref{KNmetric}). We separate the coordinates in the solutions to equation (\ref{KG1}) as
\begin{equation}
\Phi(t, r, \theta, \phi) = \mathrm{e}^{- i \omega t + i m \phi} R(r) S(\theta)\label{phi-expand1}.
\end{equation}
Using (\ref{phi-expand1}) in equation (\ref{KG1})  leads to two differential equations for angular function  $S(\theta)$ and the radial function $R(r)$,
\begin{eqnarray}
\frac1{\sin\theta} \partial_\theta (\sin\theta \, \partial_\theta S(\theta)) - \left[ a^2 \omega^2 \sin^2\theta + \frac{m^2}{\sin^2\theta} - K_l \right] S(\theta) = 0\label{angular1},
\\
\partial_r (\Delta \partial_r R(r)) + \left[ \frac{[ (r^2 + a^2) \omega - e Q r - m a ]^2}{\Delta} + 2 m a \omega - K_l \right] R(r) = 0\label{radial1},
\end{eqnarray}
 where $K_l$ is the separation constant. Furthermore, the radial equation (\ref{radial1}) can be rewritten as
\[
\partial_r (\Delta \partial_r R(r)) \!+\! \left[ \frac{\left[ (r_+^2 \!+\! a^2) \omega - a m - Q r_+ q \right]^2}{(r - r_+) (r_+ - r_-)} \!-\! \frac{\left[ (r_-^2 \!+\! a^2) \omega - a m - Q r_- q \right]^2}{(r - r_-) (r_+ - r_-)} \right] R(r)
\]
\be\label{radial}
+ f(r) R(r) = K_l R(r),
\ee
where $f(r) = \omega^2 r^2 + 2 (\omega M - e Q) \omega r +  \omega^2 a^2 - \omega^2 Q^2 + (2 \omega M - e Q)^2$.
To simplify the radial equation (\ref{radial}) and find the hidden conformal symmetry, we consider the low  frequency scalar field
$\omega M \ll 1$ where the non-extremal condition guarantees $\omega a \ll 1$ and $\omega Q \ll
1$. Moreover, we assume small electric charge for the scalar field $e Q \ll 1$. These conditions in the near region geometry $\omega
r \ll 1$, lead to neglect the function $f(r)$ in the radial equation (\ref{radial}). So, we find
\[
\partial_r (\Delta \partial_r R(r)) + \left[ \frac{\left[ (2Mr_+ - Q^2) \omega - a m - Q r_+ e \right]^2}{(r - r_+) (r_+ - r_-)} - \frac{\left[ (2Mr_- - Q^2) \omega - a m - Q r_- e \right]^2}{(r - r_-) (r_+ - r_-)} \right] R(r)
\]
\begin{equation} \label{eqR}
 = l(l+1) R(r),
\end{equation}
where we set the separation constant $K_l=l(l+1)$.
\par
Considering a charged probe in the background of a rotating charge black hole in Kerr/CFT correspondence leads to new features that are quite distinct for rotating charged black holes \cite{Chen-2fold, Ghezelbash} . 
As the first example, there are two different individual $CFT_2$ that are holographically dual to the Kerr-Newman black hole. The twofold hidden conformal symmetries are in $J$ picture where the charge of probe is assumed to be very small and $Q$ picture where the probe co-rotates with the horizon. In $J$ picture the electric charge of probe is set to be zero while in $Q$ picture the scalar wave expansion is restricted to be in the $m=0$ mode. Each of the two pictures provides the hidden conformal symmetry and so establishes the correspondence to the $CFT_2$ \cite{Chen-2fold}. 
\par
Therefore, one may expect the twofold hidden conformal symmetries must exist for other four-dimensional rotating charged black holes. 
However, the Kerr-Sen black hole which is a rotating charged black hole in four-dimensions doesn't possess the twofold hidden conformal symmetries. 
More specifically, the four-dimensional Kerr-Sen black hole as the solutions to the low energy limit of heterotic string theory don't have the hidden conformal symmetry in a well defined $Q$ picture  \cite{Ghezelbash}.
One may consider the absence of $Q$ picture for the Kerr-Sen black hole as a counterexample to the ``microscopic hair conjecture'' that only exists in Einstein-Maxwell theory  \cite{Chen-2fold}. 
\par
In revealing the twofold picture of hidden conformal symmetries for the Kerr-Newman black holes, the scalar wave function can be expanded as 
\be
\Phi = e^{-\omega t + im\phi + ie\chi} R(r) S(\theta),\label{5dwave}
\ee
where the internal dimension $\chi$ has the same $U(1)$ symmetry as the coordinate $\phi$. The existence of two coordinates with $U(1)$ symmetry leads the twofold hidden symmetries for the Kerr-Newman black holes. We note that  
the twofold hidden conformal symmetries of the Kerr-Newman suggest the unique central charge in each picture. In $J$ picture, the central charge depends only on the angular momentum $J$ while in $Q$ picture, the central charge depends only on the black holes charge $Q$. In both pictures, all the results for microscopic entropy, absorption cross section, and real time correlators are in favor of Kerr/CFT correspondence. 
In the next following sections, we confirm that the deformed hidden conformal symmetry for Kerr-Newman black hole exists in both $J$ and $Q$ pictures, as well as finally can be collected in a single picture namely general picture \cite{Chen-genhidden4d5d}.

\section{Deformed hidden conformal symmetry in $J$ picture}\label{deformedJ}
The radial equation (\ref{eqR}) has two poles on outer horizon $r_+$ and inner horizon $r_-$ where the Kerr-Newman metric function $\Delta$ (\ref{deltainkerr}) vanishes. 
For the Kerr-Newman black holes far from the extremality, we note that $r$ is far enough from $r_-$. As a result of this, we can drop the linear and quadratic terms in frequency \cite{Ghezelbash,Lowe}. These terms are coming from the expansion near the inner horizon.
So we deform the radial equation (\ref{eqR}) 
near the inner horizon $r_-$ 
by deformation parameter $\kappa$ as 
\[
\partial_r (\Delta \partial_r R(r)) + \left[ \frac{\left[ (2Mr_+ - Q^2) \omega - a m - Q r_+ e \right]^2}{(r - r_+) (r_+ - r_-)} - \frac{\left[ (2M\kappa r_+ - Q^2) \omega - a m - Q \kappa r_+ e \right]^2}{(r - r_-) (r_+ - r_-)} \right] R(r)
\]
\be \label{deformedKN}
= l (l + 1) R(r).
\ee
where $\kappa$ satisfies 
$(2M^2\kappa-Q^2)am \omega << 2\sqrt{M^2-a^2-Q^2}(r-r_-)$ as well as 
$(2M^2\kappa-Q^2)^2 \omega^2 << 2\sqrt{M^2-a^2-Q^2}(r-r_-)$. These two constraints on $\kappa$ guarantee that equation (\ref{deformedKN}) is still in low frequency limit and so one can neglect the linear and quadratic terms in frequency that come from the expansion near the inner horizon. Moreover the constraints do not change drastically the near region geometry of the black hole. We note that physical justification for the deformation in the radial equation near the inner horizon is related to the fact that the solutions to the exact radial equation (\ref{radial}) (before going to the near region and considering the low frequency limit and small electric charge for the scalar field) are singular at the inner horizon. However it is shown that the back-reaction of the field on the internal geometry of black hole replaces the inner Cauchy horizon by a null curvature spacelike singularity that covers the inner horizon of the black hole \cite{back}. As a result, the region behind the null spacelike singularity that includes the inner horizon is not the physical region of interest in the solutions of the radial equation (\ref{eqR}). In other words, one can consider the deformation of the radial equation near the inner horizon only, given by (\ref{deformedKN}), as the radial equation describes the dynamics of the test field outside of the null spacelike singularity. The other interesting feature of deformation of the inner horizon is that it doesn't change the location of other singularities of the radial equation (\ref{eqR}) that are located on the outer horizon and far infinity.  

Let us consider first the deformed equation (\ref{deformedKN}) in the $J$ picture that can be written as 
\be \label{deformedKN-J}
\partial_r (\Delta \partial_r R(r)) + \left[ \frac{\left[ (2M r_+ - Q^2) \omega - a m\right]^2}{(r - r_+) (r_+ - r_-)} - \frac{\left[ (2M\kappa r_+ - Q^2) \omega - a m\right]^2}{(r - r_-) (r_+ - r_-)} \right] R(r) = l (l + 1) R(r).
\ee
We consider the following vector fields 
\begin{eqnarray}%
  L_ \pm   &=& e^{ \pm \rho t \pm \sigma \phi } \left( { \mp \sqrt \Delta  \partial _r  + \frac{{C_1  - \gamma r}}
{{\sqrt \Delta  }}\partial _t  + \frac{{C_2  - \delta r}}
{{\sqrt \Delta  }}\partial _\phi  } \right) \label{Lpm},\\
  L_0  &=& \gamma \partial _t  + \delta \partial _\phi , \label{L0} 
\end{eqnarray}
that make the $sl(2,\mb{R})$ algebra given by $\left[ {L_ \pm  ,L_0 } \right] =  \pm L_ \pm $ and $\left[ {L_ +  ,L_ -  } \right] = 2L_0 $ \cite{Ghezelbash,Lowe}. Moreover we require the quadratic Casimir operator of $sl(2,\mb{R})$ represents the deformed radial equation (\ref{deformedKN-J}). 
Hence we find 
\begin{align}
L_0 ^2  - \frac{1}{2}\left( {L_ +  L_ -   + L_ -  L_ +  } \right) = \partial _r \left( {\Delta \partial _r } \right) + \frac{{\left( {(2Mr_ + - Q^2)  \omega  - am } \right)^2 }}{{\left( {r - r_ +  } \right)\left( {r_ +   - r_ -  } \right)}} - \frac{{\left( {(2M\kappa r_ +  - Q^2)  \omega  - am  } \right)^2 }}{{\left( {r - r_ -  } \right)\left( {r_ +   - r_ -  } \right)}}.\label{casdef}
\end{align}
We notice that the following automorphism for the generators $L_\pm$ and $L_0$,
\be \label{automorphism}
L_\pm\to -L_\pm~~~,~~~L_0 \to L_0,
\ee
does not change the $sl(2,\mathbb{R})$ algebra and so the quadratic Casimir operator is invariant.
\par 
We get the following two equations for the coefficients of $\partial _r$ and $\partial _r^2$ in (\ref{casdef}) 
\begin{align}
	\rho C_1  + \sigma C_2  + M = 0\label{rhosig1},
\end{align}
and
\begin{align}
	1 + \rho \gamma  + \sigma \delta  = 0\label{rhosig2}.
\end{align}
Moreover, the coefficients of $\partial _\phi ^2$ and $\partial _t^2$ in (\ref{casdef}) give two other equations as 
\begin{align}
	- \delta ^2 \left( {r - r_ +  } \right)\left( {r - r_ -  } \right) + C_2 ^2  - 2C_2 \delta r + \delta ^2 r^2  = a^2\label{m},
\end{align}
and
\[
	C_1 ^2  - \gamma^2 \left( {r - r_ +  } \right)\left( {r - r_ -  } \right) - 2C_1 \gamma r + \gamma ^2 r^2  = \frac{{(2Mr_+ -Q^2)^2 }}{{\left( {r_ +   - r_ -  } \right)}}\left( {\left( {r - r_ -  } \right) - \kappa ^2 \left( {r - r_ +  } \right)} \right)-
\]
\be
 - \frac{{4MQ^2 r_ +  }}
{{\left( {r_ +   - r_ -  } \right)}}\left( {r - r_ -   - \kappa \left( {r - r_ +  } \right)} \right) + Q^4 .
\label{t}
\ee
Finally, we get the following equation which is the coefficient of $\partial _\phi  \partial _t$ in
(\ref{casdef}) 
\[
	 - C_2 C_1  + \delta rC_1  - \delta r^2 \gamma  + \gamma \left( {r - r_ +  } \right)\left( {r - r_ -  } \right)\delta  + C_2 \gamma r= \]
\be
=  - \frac{{2Mr_ + a }}{{\left( {r_ +   - r_ -  } \right)}}\left( { \left( {r - r_ -  } \right) - \kappa \left( {r - r_ +  } \right)} \right) + 2aQ^2\label{mt}.
\ee
From equation (\ref{m}), we find two classes of solutions, 
\begin{eqnarray}
\delta_a^J  = \frac{{2a}}{{r_ +   - r_ -  }}, \, && \, C_{2a}^J  = \frac{{a (r_ +   + r_ -)  }}{{r_ +   - r_ -  }}\label{delC2Ja},\\
\delta_b^J  = 0, \, && \, C_{2b}^J  = a\label{delC2Jb}.
\end{eqnarray}
Substituting  (\ref{delC2Ja}) and (\ref{delC2Jb}) into equations (\ref{t}) and (\ref{mt}), we find $C_1$ and $\gamma$ that are given by
\begin{eqnarray}
\gamma_a^J  = \frac{{2Mr_ +  \left( {\kappa  + 1} \right) - 2Q^2}}{{r_ +   - r_ -  }}, \, && \, C_{1a}^J  = \frac{{2Mr_ +  \left( {\kappa r_ +   + r_ -  } \right)}}{{r_ +   - r_ -  }} - Q^2\left(\frac{r_+ + r_-}{r_+ - r_-}\right)\label{gamC1a}, \\
\gamma_b^J  = \frac{{2Mr_ +  \left( {\kappa  - 1} \right)}}{{r_ +   - r_ -  }}, \, && C_{1b}^J  = \frac{{2Mr_ +  \left( {\kappa r_ +   - r_ -  } \right)}}{{r_ +   - r_ -  }}-Q^2\label{gamC1b}.
\end{eqnarray}
Solving (\ref{rhosig1}) and (\ref{rhosig2}) for $\sigma$ and $\rho$ gives all the conformal generators  (\ref{Lpm}) and (\ref{L0}) where all the constants are given in table 1.
\begin{center}
\captionof{table}{Solutions for deformed conformal generators in $J$ picture} \label{tab:tab1} 
    \begin{tabular}{ | l | l | l | p{5cm} |}
    \hline
     & branch a & branch b \\ \hline
    $\delta$ & $\frac{{2a}}{{r_ +   - r_ -  }}$ & $0$ \\ \hline
    $\gamma$ & $\frac{{2Mr_ +  \left( {\kappa  + 1} \right) - 2Q^2}}{{r_ +   - r_ -  }}$ & $\frac{{2Mr_ +  \left( {\kappa  - 1} \right)}}{{r_ +   - r_ -  }}$ \\ \hline
    $C_1$ & $ \frac{{2Mr_ +  \left( {\kappa r_ +   + r_ -  } \right)}}{{r_ +   - r_ -  }} - Q^2\left(\frac{r_+ + r_-}{r_+ - r_-}\right)$  &  $\frac{{2Mr_ +  \left( {\kappa r_ +   - r_ -  } \right)}}{{r_ +   - r_ -  }}-Q^2$ \\
    \hline
$C_2$ & $\frac{{a (r_ +   + r_ -)  }}{{r_ +   - r_ -  }}$  & $a$   \\ \hline
$\rho$ & $0$  & $-\frac{r_+ - r_-}{2(\kappa-1)Mr_+}$   \\ \hline
$\sigma$ & $\frac{{(r_ -   - r_ +)  }}{2a}$  & $\frac{{2Mr_ +  \left( {\kappa r_ +   - r_ -   - M\left( {\kappa  - 1} \right)} \right) - Q^2 \left( {r_ +   - r_ -  } \right)}}{{2aMr_ +  \left( {\kappa  - 1} \right)}}$   \\ \hline
    \end{tabular}
\end{center}
We note that multiplying all the coefficients in table 1 with $-1$ also are solutions to equations (\ref{rhosig1}), (\ref{rhosig2}), (\ref{m}), (\ref{t}) and (\ref{mt}). However these solutions correspond to invariance of the Casmir operator 
$ L_0^2 - \frac{1}{2}(L_+L_-+L_-L_+)
$
by renaming the vector fields as
\be
L_0  \to  - L_0 ~~,~~
L_ \pm   \to  - L \mp .\label{renameLs}
\ee
We also note that in the limit of $Q=0$, the vector fields in the $J$ picture reduce correctly to the generators of deformed conformal symmetry for the Kerr black holes \cite{Lowe}.
\par

Furnished by the explicit expressions for the deformed conformal generators in branch a
\begin{eqnarray}
L_{\pm}^a & = & \begin{array}{r}
e^{\mp2\pi T_{R}\phi}\left[\mp\sqrt{\Delta}\partial_{r}-\frac{1}{2\pi T_{H}}\frac{r-M}{\sqrt{\Delta}}\left(\Omega_H\partial_{\phi}+\partial_{t}\right)
+\frac{1}{2\pi\Omega_H(T_{L}+T_{R})}\frac{r-r_{+}}{\sqrt{\Delta}}\partial_{t}\right],\\
\end{array}\label{eq:L-a}\\
L_{0}^a & = & \frac{1}{2\pi T_{H}}\left(\Omega_H\partial_{\phi}+\partial_{t}\right)-\frac{1}{2\pi\Omega_H(T_{L}+T_{R})}\partial_{t},
\nonumber \end{eqnarray}
and branch b
\begin{eqnarray}
{L}_{\pm}^b & = & \begin{array}{r}
e^{\pm2\pi\Omega(T_{L}+T_{R})t\mp2\pi T_{L}\phi}\left[\mp\sqrt{\Delta}\partial_{r}+\frac{2Mr_{+}-Q^2}{\sqrt{\Delta}}\left(\Omega\partial_{\phi}+\partial_{t}\right)%\right.\\
%\left.
+\frac{1}{2\pi\Omega_H(T_{L}+T_{R})}\frac{r-r_{+}}{\sqrt{\Delta}}\partial_{t}\right],
\end{array}\label{eq:Lbar-a}\nonumber\\
&&\\
{L}_{0}^b& = & -\frac{1}{2\pi\Omega_H(T_{L}+T_{R})}\partial_{t},\nonumber 
\end{eqnarray}
where $T_H$ and $\Omega_H$ are defined in (\ref{en2}) and (\ref{en3}) and 
the left and right moving CFT temperatures are given by 
\be
T_R = \frac{r_+ - r_-}{4\pi a} ,~~~ T_L = \frac{T_R(1+\kappa)}{1-\kappa} - \frac{Q^2 T_R}{Mr_+ (1-\kappa)}\label{TLRJ}.
\ee
One can verify that taking the left and right central charges  
\be
c_R = c_L = \frac{6aMr_+ (1-\kappa)}{\sqrt{M^2-a^2-Q^2}}\label{cJ},
\ee
leads to the exact Bekenstein-Hawking entropy for Kerr-Newman black holes (\ref{en1}), if we use the Cardy formula
\be
S_{Cardy} = \frac{\pi^2}{3}(c_R T_R + c_L T_L)\label{SJ}.
\ee
We notice that for the special case of deformation parameter given by 
$\kappa = r_+/r_-$, we find the generators of hidden conformal symmetry for the Kerr-Newman black holes \cite{Chen-2fold}. %by Chen et al. 
In fact, for $\kappa = r_+/r_-$, the deformed generators $L_{\pm}^a$ and $L_0^a$ (up to automorphisms (\ref{automorphism})) reduce to conformal generators $H_\pm$ and $H_0$ in \cite{Chen-2fold} according to
\be \label{mapA}
L_k^a  =  - iH_k ,
\ee 
where $k = +,-,0$. The generators in branch b for $\kappa = r_+/r_-$ reduce to the other copy of conformal generators $\bar H_k$ in \cite{Chen-2fold} by the mapping
\be\label{mapB} 
L_k^b  = i\bar H_k .
\ee
The left and right temperatures (\ref{TLRJ}) as well as central charge (\ref{cJ}) reduce to the corresponding results in \cite{Chen-2fold} after setting $\kappa = r_-/r_+$.
\par 

An interesting open question is to derive the deformed central charges by using either ASG or stretched horizon techniques. 
\section{Deformed hidden conformal symmetry in $Q$ picture}\label{deformedQ}
The deformed radial equation (\ref{deformedKN}) in the
$Q$ picture is 
\be \label{deformedKN-Q}
\partial_r (\Delta \partial_r R(r)) + \left[ \frac{\left[ (2Mr_+ - Q^2) \omega - Q r_+ e \right]^2}{(r - r_+) (r_+ - r_-)} - \frac{\left[ (2M\kappa r_+ - Q^2) \omega - Q \kappa r_+ e \right]^2}{(r - r_-) (r_+ - r_-)} \right] R(r) = l (l + 1) R(r).
\ee 
Matching the Casimir operator of $sl(2,{\mb R})$ algebra to the left hand side of equation (\ref{deformedKN-Q})
gives the same equations (\ref{rhosig1}) and (\ref{rhosig2}) for the coefficients of  $\partial_r$ and $\partial_r^2$ in $J$ picture.  However the other equations are different and their solutions again provide two branches. The solutions are represented in table 2.
\begin{center}
\captionof{table}{Solutions for conformal generators in $Q$ picture} \label{tab:tab3} 
    \begin{tabular}{ | l | l | l | p{5cm} |}
    \hline
     & branch a & branch b \\ \hline
    $\delta$ & $\frac{{Qr_+ (1+\kappa)}}{{r_ +   - r_ -  }}$ & $\frac{{Qr_+ (\kappa-1)}}{{r_ +   - r_ -  }}$ \\ \hline
    $\gamma$ & $\frac{{2Mr_ +  \left( {\kappa  + 1} \right) - 2Q^2}}{{r_ +   - r_ -  }}$ & $\frac{{2Mr_ +  \left( {\kappa  - 1} \right)}}{{r_ +   - r_ -  }}$ \\ \hline
    $C_1$ & $ \frac{{2Mr_ +  \left( {\kappa r_ +   + r_ -  } \right)}}{{r_ +   - r_ -  }} - Q^2\left(\frac{r_+ + r_-}{r_+ - r_-}\right)$  &  $\frac{{2Mr_ +  \left( {\kappa r_ +   - r_ -  } \right)}}{{r_ +   - r_ -  }}-Q^2$ \\
    \hline
$C_2$ & $\frac{{Qr_+ (\kappa r_ +   + r_ -)  }}{{r_ +   - r_ -  }}$  & $\frac{{Qr_+ (\kappa r_ +   - r_ -)  }}{{r_ +   - r_ -  }}$  \\ \hline
$\rho$ & $\frac{r_+ - r_-}{2Q^2}$  & $\frac{M(\kappa-1)-\kappa r_+ + r_-}{Q^2(\kappa-1)}$  \\ \hline
$\sigma$ & $-\frac{M(r_+ - r_-)}{Q^3}$  & $\frac{(r_+ - r_-)(Mr_+(\kappa+1)-Q^2)}{r_+(\kappa-1)Q^3}$  \\ \hline
    \end{tabular}
\end{center}

Moreover we note that multiplying all the solutions in table 2 by $-1$ also satisfy the full set of equations. However similar to the $J$ picture, these solutions  correspond to invariance of the Casimir operator under renaming (\ref{renameLs}). 
Considering the right and left temperatures to be proportional to $\sigma$ in branches a and b as
\be
T_R = \frac{M(r_+ - r_-)}{2\pi Q^3} ~~~,~~~ T_L = T_R\frac{(1+\kappa)}{(1-\kappa)} -\frac{T_R Q^2}{Mr_+(1-\kappa)}\label{TLRQ},
\ee
one can produce the correct Bekenstein-Hawking entropy of the Kerr-Newman black holes using the Cardy formula by the central charges
\be\label{cQ}
c_L = c_R = \frac{3Q^3r_+(1-\kappa)}{\sqrt{M^2-a^2-Q^2}}.
\ee

We note that the dependence of $T_L$ in (\ref{TLRQ}) to $T_R$ has exactly the same form as the left temperature in $J$ picture (\ref{TLRJ}).  We notice that for special value of $\kappa = r_-/r_+$, the mappings (\ref{mapA}) and (\ref{mapB}) show that deformed conformal generators correctly reduce to conformal generators of the Kerr-Newman black hole.
The temperatures (\ref{TLRQ}) and the central charges (\ref{cQ}) reduce to the left and right temperatures and the central charge of CFT dual to Kerr-Newman black hole \cite{Chen-2fold}.  We note that in 
$Q$ picture in which the coefficients of conformal generators are given in table \ref{tab:tab3}, the deformed hidden conformal symmetry generators are (\ref{Lpm}) and (\ref{L0}), replacing the coordinate $\phi$ with the internal coordinate $\chi$.
%%%%%%%%%%%%%%%%%%%%%%%%%%
\section{Deformed hidden conformal symmetry  in general picture}\label{deformedGen}
As it was mentioned in introduction, the Kerr-Newman black holes have two conformal pictures as $\phi '$ and $\chi '$ pictures. These pictures correspond respectively to two separated $U(1)$ symmetries with respect to coordinates $\phi$ and $\chi$. The third conformal picture (general picture) can be obtained by using the 
modular group $SL(2,\mathbb{Z})$ of the torus  $(\phi,\chi)$. In this picture, the $SL(2,\mathbb{Z})$ transformation for the torus is given by \cite{chen-novelsl2z,Chen-genhidden4d5d}
\begin{equation}
\left( {\begin{array}{*{20}c}
   {\phi '}  \\
   {\chi '}  \\
\end{array}} \right) = \left( {\begin{array}{*{20}c}
   \alpha  & \beta   \\
   \eta  & \tau   \\
\end{array}} \right)\left( {\begin{array}{*{20}c}
   \phi   \\
   \chi   \\
\end{array}} \right),\label{SL2trans}
\end{equation} 
where $
\left( {\begin{array}{*{20}c}
   \alpha  & \beta   \\
   \eta  & \tau   \\

 \end{array} } \right) 
$
is any $SL(2,\mathbb{Z})$ group element. Under transformation (\ref{SL2trans}), the phase factor of the charged scalar field (\ref{phi-expand1})  with the electric charge $e$ is invariant;
$e^{im\phi  + ie\chi }  = e^{im'\phi ' + ie'\chi '} $ which yields 
\be
m = \alpha m' + \eta e'~~~,~~~e = \beta m' + \tau e'.\label{em-em}
\ee
In $\phi'$ picture, we set $e' = 0$, hence the deformed radial equation  is
\bea
&&\partial _r \left( {\Delta \partial _r R\left( r \right)} \right)  \nonumber\\
&+&\left( {\frac{{\left( {(2Mr_ + - Q^2 ) \omega  - \left( {Qr_ +  \beta + a\alpha} \right)m' } \right)^2 }}{{\left( {r - r_ +  } \right)\left( {r_ +   - r_ -  } \right)}}} %\right.
%\ee
%\be
 - %\left.
 {\frac{{\left( {(2M\kappa r_ + - Q^2 ) \omega  - \left( {Q\kappa r_ +  \beta + a\alpha} \right)m' } \right)^2 }}{{\left( {r - r_ -  } \right)\left( {r_ +   - r_ -  } \right)}}} \right)R\left( r \right) \nonumber\\
&=&l\left( {l + 1} \right)R\left( r \right)\label{deformedKNsl2z},
\eea
Similar to $J$ and $Q$ pictures, we match the Casimir operator of $sl(2,{\mb R})$ to the left hand side of equation (\ref{deformedKNsl2z}) and solve for the coefficients of the vector fields (\ref{Lpm}) and (\ref{L0}).
We find there are two classes of solutions for $\delta,\gamma,C_1$, and $C_2$ that are given by
\begin{eqnarray}
\delta_a^G  = \frac{a_1 + a_2}{{r_ +   - r_ -  }}, \, && \, C_{2a}^G  = \frac{a_1 r_- + a_2 r_+}{{r_ +   - r_ -  }},\\
\delta_b^G  = \frac{a_2 - a_1}{{r_ +   - r_ -  }}, \, && \, C_{2b}^G  = \frac{a_2 r_+ - a_1 r_-}{{r_ +   - r_ -  }},
\end{eqnarray}\label{delC2sl2z}
\begin{eqnarray}
\gamma_a^G  = \frac{{2Mr_ +  \left( {\kappa  + 1} \right) - 2Q^2}}{{r_ +   - r_ -  }}, \, && \, C_{1a}^G  = \frac{{2Mr_ +  \left( {\kappa r_ +   + r_ -  } \right)}}{{r_ +   - r_ -  }} - Q^2\left(\frac{r_+ + r_-}{r_+ - r_-}\right), \\
\gamma_b^G  = \frac{{2Mr_ +  \left( {\kappa  - 1} \right)}}{{r_ +   - r_ -  }}, \, && C_{1b}^G  = \frac{{2Mr_ +  \left( {\kappa r_ +   - r_ -  } \right)}}{{r_ +   - r_ -  }}-Q^2,
\end{eqnarray}\label{gamC1sl2z}
where
\be
a_1 = Qr_+ \beta + a\alpha ~~~,~~~ a_2 = Q\kappa r_+ \beta + a\alpha \label{a1a2}.
\ee
Table 3 shows the full set of solutions for branch a and b.

\begin{center}
\captionof{table}{Solutions for conformal generators in general picture} \label{tab:tab4} 
    \begin{tabular}{ | l | l | l | p{5cm} |}
    \hline
     & branch a & branch b \\ \hline
    $\delta$ & $\frac{2\alpha a + \left( {\kappa  + 1} \right)\beta Qr_ +}{{r_ +   - r_ -  }}$ & $\frac{\left( {\kappa  - 1} \right)\beta Qr_ +}{{r_ +   - r_ -  }}$ \\ \hline
    $\gamma$ & $\frac{{2Mr_ +  \left( {\kappa  + 1} \right) - 2Q^2}}{{r_ +   - r_ -  }}$ & $\frac{{2Mr_ +  \left( {\kappa  - 1} \right)}}{{r_ +   - r_ -  }}$ \\ \hline
    $C_1$ & $\frac{{2Mr_ +  \left( {\kappa r_ +   + r_ -  } \right)}}{{r_ +   - r_ -  }} - Q^2\left(\frac{r_+ + r_-}{r_+ - r_-}\right)$  &  $\frac{{2Mr_ +  \left( {\kappa r_ +   - r_ -  } \right)}}{{r_ +   - r_ -  }}-Q^2$ \\
    \hline
$C_2$ & $\frac{Q\beta r_+ (\kappa r_+ +r_-)+a\alpha (r_+ + r_-)}{{r_ +   - r_ -  }}$  & $\frac{Q\beta r_+ (\kappa r_+ -r_-)+a\alpha (r_+ - r_-)}{{r_ +   - r_ -  }}$  \\ \hline
$\rho$ & $\frac{Q\beta (r_+ - r_-)}{2 (2Ma\alpha +Q^3 \beta)}$  & $ - \frac{{\alpha a(r_ +   - r_ -  ) - Q\beta r_ +  (M(\kappa  - 1) - \kappa r_ +   + r_ -  )}}
{{(\kappa  - 1)r_ +  (2Ma\alpha  + Q^3 \beta )}}$  \\ \hline
$\sigma$ & $-\frac{M (r_+ - r_-)}{(2Ma\alpha +Q^3 \beta)}$  & $\frac{{(r_ +   - r_ -  )((\kappa  + 1)Mr_ +   - Q^2 )}}
{{(\kappa  - 1)r_ +  (2Ma\alpha  + Q^3 \beta )}}$  \\ \hline
    \end{tabular}
\end{center}
\par
In this picture, the left and right CFT temperatures are given by
\be
T_R = \frac{M(r_+ - r_-)}{2\pi(2Ma\alpha + Q^3 \beta)}~~~,~~~ T_L = T_R\frac{(1+\kappa)}{(1-\kappa)} - \frac{T_R Q^2}{Mr_+(1-\kappa)}\label{TLRgen}.
\ee
The agreement between microscopic CFT entropy and the Hawking-Bekenstein entropy requires that the central charges are 
\be
c_L = c_R = \frac{3(1-\kappa)r_+(2Ma\alpha + Q\beta)}{\sqrt{M^2-a^2-Q^2}}.
\ee
We note that the right temperature of generalized CFT is independent of deformation parameter $\kappa$. However the left temperature non-trivially depends on the deformation parameter $\kappa$. 
Moreover, we should note that the deformed hidden conformal symmetry generators in general $\phi '$ picture are given by (\ref{Lpm}) and (\ref{L0}), replacing the coordinate $\phi$ by $\phi '$.
The solutions for coefficients (tabulated in table \ref{tab:tab4}) reduce to the corresponding coefficients in table \ref{tab:tab1}  in $J$ picture  where we set $\alpha = 1,\beta=0$, and reduce to the coefficients in table \ref{tab:tab3} in $Q$ picture where $\alpha = 0,\beta = 1$. As a result the generators in the $\phi '$ picture reduce to the corresponding generators in $J$ and $Q$ pictures respectively.

\section{Scattering of charged scalars in the Kerr-Newman background based on deformed radial equation}\label{scat}
In this section, we consider the absorption cross section of the scalar fields in the background of Kerr-Newman black holes in different pictures. 

\subsection{$J$ picture}

We re-write the deformed equation in $J$ picture (\ref{deformedKN}) as
\be
\partial _r \left( {\Delta \partial _r R\left( r \right)} \right) + \left( {\frac{{(g_ + ^J)^2 \left( {r_ +   - r_ -  } \right)}}
{{\left( {r - r_ +  } \right)}} - \frac{{(g_ - ^J)^2 \left( {r_ +   - r_ -  } \right)}}
{{\left( {r - r_ -  } \right)}} - K_l } \right)R\left( r \right) = 0\label{defeomKNJ}
\ee
where
\begin{eqnarray}
\label{newvarKNeom-J}
  g_ + ^J  &=& \frac{\left( {2Mr_ +   - Q^2 } \right)\omega  - am}{r_+ - r_-}, \\
  g_ - ^J  &=& \frac{\left( {2M\kappa r_ +   - Q^2 } \right)\omega  - am}{r_+ - r_-}. 
\end{eqnarray} 

We define the new coordinate \cite{Malda-Strom}
\begin{align}
p = \frac{{r - r_ +  }}{{r - r_ -  }}\label{p},
\end{align}
and so the deformed equation  (\ref{defeomKNJ}) becomes 
\begin{align}
p\left( {1 - p} \right)\partial _p^2 R\left( p \right) + \left( {1 - p} \right)\partial _p R\left( p \right) + \left( {\frac{{g_+^J }}{p} - g_-^J  - \frac{{K_l }}{{1 - p}}} \right)R\left( p \right) = 0,\label{eq-p}
\end{align}
where we used the following identity% (\ref{p}),
\begin{align}
\Delta \partial _r  = \left( {r_ +   - r_ -  } \right)p\partial _p.\label{pp}
\end{align}
The in-going solution for the equation (\ref{eq-p}) is
\begin{align}
	R_{in} \left( r \right) = Const. {\rm{ }}p^{ - ig_ + ^J } \left( {p - 1} \right)^{ - l} {}_2F_1 \left( { - l - i\left( {g_ + ^J  - g_ - ^J } \right), - l - i\left( {g_ + ^J  + g_ - ^J } \right);1 - 2ig_ + ^J ;p} \right),\label{Rin}
\end{align}
where $ {}_2 F_1 $ is the hypergeometric function. The in-going solution (\ref{Rin}) on the outer boundary of the matching region where $r>>M$ behaves as, 
\begin{align}
R_{in}  \sim Ar^l,\label{Rin-far}
\end{align}
where $ 
A = {}_2F_1 \left( { - l - i\left( {g_ + ^J  - g_ - ^J } \right), - l - i\left( {g_ + ^J  + g_ - ^J } \right);1 - 2ig_ + ^J ;1} \right)
$. We should mention in finding the in-going solution, we consider the low frequency condition, $\omega  <  < 1/M$ in near region, $r <  < 1/\omega$. 
Using the Gauss' theorem for hypergeometric functions, we can re-write the factor $A$ in equation (\ref{Rin-far}) as
\begin{align}
	A = \frac{{\Gamma \left( {1 - 2ig_ + ^J } \right)\Gamma \left( {2l + 1} \right)}}{{\Gamma \left( l + 1 - 2i\left( {\frac{{\left( {Mr_ +  \left( {1 + \kappa } \right) - Q^2 } \right)\omega  - am}}
{{r_ +   - r_ -  }}} \right) \right)\Gamma \left( l + 1 - 2i\left( {\frac{{Mr_ +  \left( {1 - \kappa } \right)\omega }}
{{r_ +   - r_ -  }}} \right) \right)}}.\label{AGauss}
\end{align}
Hence, we find the absorption cross section, given by 
\begin{align}
	P_{abs}  \sim \left| A \right|^{-2}  = \sinh \left( {2\pi g_+^J } \right)\frac{{\left| {\Gamma \left( {l + 1 - iB_1 } \right)} \right|^2 \left| {\Gamma \left( {l + 1 - iB_2 } \right)} \right|^2 }}{{2\pi g_+^J \left( {\Gamma \left( {2l + 1} \right)} \right)^2 }}\label{PabsJ},
\end{align}
where
\begin{eqnarray}\label{B1B2J}
B_1  &=& 2\left( {\frac{{\left( {Mr_ +  \left( {1 + \kappa } \right) - Q^2 } \right)\omega  - am}}
{{r_ +   - r_ -  }}} \right),\\
B_2  &=&2\left( {\frac{{Mr_ +  \left( {1 - \kappa } \right)\omega }}
{{r_ +   - r_ -  }}} \right).
\end{eqnarray}
In supporting the Kerr/CFT duality in this scattering process, 
we need to associate the absorption cross section (\ref{PabsJ}) with the results from 2D CFT. In other words, we want to match the absorption cross section (\ref{PabsJ}) computed from gravitational side to the corresponding cross section in the dual 2D CFT in $J$ picture, 
\begin{align}
	P_{abs}  \sim {T^J_L}^{2h_L  - 1} {T^J_R}^{2h_R  - 1} \sinh \left( {\frac{{\omega^J _L }}{{2{T^J_L} }} + \frac{{\omega^J _R }}{{2{T^J_R} }}} \right)\left| {\Gamma \left( {h_L  + i\frac{{\omega^J _L }}{{2\pi {T^J_L} }}} \right)} \right|^2 \left| {\Gamma \left( {h_R  + i\frac{{\omega^J _R }}{{2\pi {T^J_R} }}} \right)} \right|^2 \label{Pabs2CFTJ}
\end{align}
which is known as the finite temperature absorption cross section in a 2D CFT \cite{Malda-Strom}. To match and find the possible agreement between (\ref{Pabs2CFTJ}) and (\ref{PabsJ}), we consider the first law of  thermodynamics for the charged rotating black holes
\be \label{BHthermoLaw}
T_H \delta S_{BH}  = \delta M - \Omega _H \delta J - \Phi _H \delta Q.
\ee 
where $T_H$,$\Omega_H$ and $\Phi_H$ are given by (\ref{en2}), (\ref{en3}) and (\ref{en4}). For a 2D CFT with the Cardy entropy 
\cite{Compere}
\be 
S_{CFT}  = 2\pi \left( {\sqrt {\frac{{c_L E_L }}{6}}  + \sqrt {\frac{{c_R E_R }}{6}} } \right),
\ee 
the variation of entropy can be read as
\be \label{delSCFT}
\delta S_{CFT}  = \frac{{\delta E_L }}{{T_L }} + \frac{{\delta E_R }}{{T_R }}.
\ee 
Matching the variations of entropy (\ref{BHthermoLaw}) and CFT entropy (\ref{delSCFT}) gives 
\be  \label{Entropy-matchingJ}
\frac{\delta M - \Omega _H \delta J - \Phi _H \delta Q}{T_H} = \frac{{\delta E^J_L }}{{T^J_L }} + \frac{{\delta E^J_R }}{{T^J_R }}.
\ee
In the last equation and also in (\ref{Pabs2CFTJ}), the superscripts $J$ show the corresponding quantities in the $J$ picture. We can identify $\delta M$ as $\omega$, $\delta J$ as $m$, $\delta Q$ as $e$, $\delta E^J_{R,L} = \omega^J_{R,L}$ in (\ref{Entropy-matchingJ}). 
Therefore a set of left and right frequencies that satisfy the equation (\ref{Entropy-matchingJ}) are
\be \label{LRfreqJ}
\omega _L^J  = \frac{{\omega \left( {Mr_ +  \left( {\kappa  + 1} \right) - Q^2 } \right)}}{a}~~,~~\omega _R^J  = \omega _L^J  - m.
\ee 
For $\kappa = r_-/r_+$, these left and right frequencies definitely reduce to the left and right frequencies in $J$ picture for the Kerr-Newman black hole \cite{Chen-2fold}.
The fact which supports the existence of dual 2D CFT for the deformed Kerr-Newman/CFT correspondence is the agreement between (\ref{Pabs2CFTJ}) with (\ref{PabsJ}) if the $\omega_{L,R}^J$ are as in (\ref{LRfreqJ}). In the formula (\ref{Pabs2CFTJ}), the left and right conformal weights $h_{L,R}$ are equal to $l+1$. We notice that these conformal weights are the same in the other $Q$ and general pictures that we discuss in next two subsections. 

\subsection{$Q$ picture}
In  $Q$ picture, the charged test particle is rotating corotationally with the black hole horizon, thus we can turn off the rotational parameter $a$.
The absorption cross section and the deformed radial equation are given by  (\ref{PabsJ}) and  
(\ref{defeomKNJ}) with replacing $g^J$ to $g^Q$ where
\begin{eqnarray}\label{newvarKNeom-Q}
  g_ + ^Q  &=& \frac{\left( {2Mr_ +   - Q^2 } \right)\omega  - Qr_+e}{r_+ - r_-},  \\
  g_ - ^Q  &=& \frac{\left( {2M\kappa r_ +   - Q^2 } \right)\omega  - Q\kappa r_+e}{r_+ - r_-} .
\end{eqnarray} 
Thus we find the corresponding absorption cross section, given by 
\begin{align}
	P_{abs}  \sim \left| A \right|^{-2}  = \sinh \left( {2\pi g_+^J } \right)\frac{{\left| {\Gamma \left( {l + 1 - iB^Q_1 } \right)} \right|^2 \left| {\Gamma \left( {l + 1 - iB^Q_2 } \right)} \right|^2 }}{{2\pi g_+^J \left( {\Gamma \left( {2l + 1} \right)} \right)^2 }}\label{PabsQ},
\end{align}
where
\begin{eqnarray}
B^Q_1  &=& \left( {\frac{{\left( {2Mr_ +  \left( {1 + \kappa } \right) - 2Q^2 } \right)\omega  - Q(1+\kappa)r_+ e}}
{{r_ +   - r_ -  }}} \right),\\
B^Q_2  &=&\left( {\frac{{2Mr_ +  \left( {1 - \kappa } \right)\omega -Q(1-\kappa)r_+ e}}
{{r_ +   - r_ -  }}} \right).
\end{eqnarray}
In this picture, to match and find the possible agreement between the cross section  (\ref{PabsQ}) and the finite temperature absorption cross section of CFT, we again consider the first law of  thermodynamics for the charged rotating black holes. 
The matching of microscopic and macroscopic entropy variations in $Q$ picture now can be read as
\be  \label{Entropy-matchingQ}
\frac{\delta M - \Omega _H \delta J - \Phi _H \delta Q}{T_H} = \frac{{\delta E^Q_L }}{{T^Q_L }} + \frac{{\delta E^Q_R }}{{T^Q_R }}.
\ee
We identify $\delta E_{L,R}$
with the left and right frequencies $\tilde{\omega}_{L,R}^Q$ that are related to three quantities; the left and right frequencies $\omega_{L,R}^Q$, charges $q_{L,R}$, and chemical potentials $\mu_{L,R}$ where 
\be \label{freqQ}
\omega^Q _R  = \omega^Q _L  = \frac{{2M\omega }}{{Q^3 }}\left( {Mr_ +  \left( {1 + \kappa } \right) - Q^2 } \right),
\ee 
\be \label{chem-potentialQ}
\mu _R  = \mu _L  + 1 = \frac{{Mr_ +  \left( {1 + \kappa } \right)}}{{Q^2 }}.
\ee 
The frequencies $\tilde \omega_{L,R}$ are given by
\be \label{gen-freq-Q}
{\tilde \omega}^Q _{L,R}  = \omega^Q _{L,R}  - q_{L,R} \mu _{L,R} ,
\ee 
where the charges $q_{L,R} = e$. Substituting equations (\ref{freqQ}), (\ref{chem-potentialQ}) and (\ref{gen-freq-Q}) into the 2D CFT absorption cross section 
\begin{align}
	P_{abs}  \sim {T^Q_L} ^{2h_L  - 1} {T^Q_R}^{2h_R  - 1} \sinh \left( {\frac{{{\tilde \omega}^Q _L }}{{2T_L }} + \frac{{{\tilde \omega}^Q _R }}{{2T^Q_R }}} \right)\left| {\Gamma \left( {h_L  + i\frac{{{\tilde \omega}^Q _L }}{{2\pi T^Q_L }}} \right)} \right|^2 \left| {\Gamma \left( {h_R  + i\frac{{{\tilde \omega}^Q _R }}{{2\pi T^Q_R }}} \right)} \right|^2 \label{Pabs2CFTQ},
\end{align}
shows that the 2D CFT cross section agrees with the absorption cross section which is derived from gravitational point of view in $Q$ picture (\ref{PabsQ}). Also as it is expected, when $\kappa = r_-/r_+$, the left and right frequencies for the deformed 2D CFT as well as the chemical potential  reduce to the corresponding quantities in \cite{Chen-2fold}.

\subsection{General picture}
The $SL(2,\mathbb{Z})$ transformation between $(\phi,\chi)$ and $(\phi ',\chi ')$ yields the relations (\ref{em-em}). The $\phi '$ picture under consideration is given by setting $e' = 0$.
The deformed radial equation (\ref{deformedKNsl2z}) in $\phi'$ picture can be rewritten as,
\be
\partial _r \left( {\Delta \partial _r R\left( r \right)} \right) + \left( {\frac{{(g_ + ^G)^2 \left( {r_ +   - r_ -  } \right)}}
{{\left( {r - r_ +  } \right)}} - \frac{{(g_ - ^G)^2 \left( {r_ +   - r_ -  } \right)}}
{{\left( {r - r_ -  } \right)}} - K_l } \right)R\left( r \right) = 0,\label{defeomKNSL2Z1}
\ee
where  $g_+^G$ and $g_-^G$ are
\begin{eqnarray}
  g_ + ^G  &=& \frac{\left( {2Mr_ +   - Q^2 } \right)\omega  - (Qr_+\beta + a\alpha)m'}{r_+ - r_-} , \\
  g_ - ^G  &=& \frac{\left( {2M\kappa r_ +   - Q^2 } \right)\omega  - (Q\kappa r_+\beta + a\alpha)m'}{r_+ - r_-} .
\end{eqnarray} 
The absorption cross section is given by 
\begin{align}
	P_{abs}  \sim \left| A \right|^{-2}  = \sinh \left( {2\pi g_+^G } \right)\frac{{\left| {\Gamma \left( {l + 1 - iB^G_1 } \right)} \right|^2 \left| {\Gamma \left( {l + 1 - iB^G_2 } \right)} \right|^2 }}{{2\pi g_+^G \left( {\Gamma \left( {2l + 1} \right)} \right)^2 }}\label{PabsSL2Z},
\end{align}
where
\begin{eqnarray}
B^G_1  &=& \left( {\frac{{\left( {2Mr_ +  \left( {1 + \kappa } \right) - 2Q^2 } \right)\omega  - (Q(1+\kappa)r_+ \beta + 2a\alpha)m'}}
{{r_ +   - r_ -  }}} \right),\\
B^G_2  &=&\left( {\frac{{2Mr_ +  \left( {1 - \kappa } \right)\omega -Q(1-\kappa)r_+ \beta m'}}
{{r_ +   - r_ -  }}} \right).
\end{eqnarray}

 Matching the macroscopic and microscopic entropy requires that we introduce the generalized frequencies $\tilde \omega^G_{L,R}$ in terms of three quantities; frequencies $\omega^G_{L,R}$, charges $q^G_{L,R}$, and chemical potentials $\mu^G_{L,R}$,
\be 
\tilde \omega _{L,R}^G  = \omega _{L,R}^G  - q_{L,R}^G \mu _{L,R}^G.\label{ff}
\ee
In (\ref{ff}), 
\be 
\omega _{L,R}^G  = \frac{{2M\omega \left( {Mr_ +  \left( {1 + \kappa } \right) - Q^2 } \right)}}{{2\alpha Ma + \beta Q^3 }},
\ee
 \be 
\mu _R^G  = \frac{{M\left( {2\alpha a + \beta Qr_ +  \left( {1 + \kappa } \right)} \right)}}{{2\alpha Ma + \beta Q^3 }}~~,~~\mu _L^G  = \frac{{\beta Q\left( {Mr_ +  \left( {1 + \kappa } \right) - Q^2 } \right)}}{{2\alpha Ma + \beta Q^3 }},
\ee 
and $q_{L,R}^G = m'$. Substituting the relevant quantities in the CFT absorption cross section
\begin{align}
	P_{abs}  \sim {T^G_L} ^{2h_L  - 1} {T^G_R}^{2h_R  - 1} \sinh \left( {\frac{{{\tilde \omega}^G _L }}{{2T_L }} + \frac{{{\tilde \omega}^G _R }}{{2T^G_R }}} \right)\left| {\Gamma \left( {h_L  + i\frac{{{\tilde \omega}^G _L }}{{2\pi T^G_L }}} \right)} \right|^2 \left| {\Gamma \left( {h_R  + i\frac{{{\tilde \omega}^G _R }}{{2\pi T^G_R }}} \right)} \right|^2 \label{Pabs2CFTgen}
\end{align}
shows exact agreement between the CFT absorption cross section (\ref{Pabs2CFTgen}) and the corresponding cross section from gravitational side (\ref{PabsSL2Z}). 
\par 

\section{The (deformed) hidden conformal symmetries of Kerr and \rn }\label{reis}

We notice that  the deformed conformal generators $L^{a,b}_n$, $n = +,-,0$ in (\ref{eq:L-a}) and (\ref{eq:Lbar-a}) reduce to the deformed conformal generators  $L_n$, $n = +,-,0$ for the Kerr black holes when we set $Q=0$ \cite{Lowe} . Moreover, setting the rotation parameter $a=0$ with special value of deformation parameter $\kappa$, we find the conformal generators of \sch black holes in agreement with \cite{Lowe} and \cite{Bertini}.
\par

Plugging the results for  $C_1,C_2,\delta,\gamma,\rho$ and $\sigma$ for branch a from Table 2 in $Q$ picture along with $a=0$, we find the 
deformed hidden conformal generators for \rn black holes as
\begin{eqnarray}
 L_ \pm ^a  &=& e^{ \pm \left( {\frac{{\pi QT_R }}{M}} \right)t \mp \left( {2\pi T_R } \right)\chi } \left( { \mp \sqrt \Delta  \partial _r  + \left( {\frac{M}{{2\pi r_ +  }}\left( {\kappa r_ +   + r_ -   - r\left( {1 + \kappa } \right)} \right) + \frac{{Q^2 }}{{4\pi r_ + ^2 }}\left( {2r - r_ +   - r_ -  } \right)} \right)\partial _t } \right.\nonumber \\ 
&&\left. { + \frac{{r_ +  }}{{Q^2 2\pi T_R }}\left( {\kappa r_ +   - r_ -   - r\left( {1 + \kappa } \right)} \right)\partial _\chi  } \right), \label{lpmrna}\\ 
L_0^a  &=& \frac{{2M}}{{2\pi Q^3 T_R }}\left( {Mr_ +  \left( {\kappa  + 1} \right) - Q^2 } \right)\partial _t  + \frac{{Mr_ +  \left( {\kappa  + 1} \right)}}{{2\pi Q^2 T_R }}\partial _\chi ~ .\label{l0rna}
 \end{eqnarray}
%\ee 
The right temperature $T_R$ is given by
\be \label{TLRN-A}
T_R  = \frac{{\left( {r_ +   - r_ -  } \right)M}}{{2\pi Q^3 }},
\ee  
where the outer and inner \rn black holes horizons are $r_+ = M+\sqrt{M^2-Q^2}$ and $r_- = M-\sqrt{M^2-Q^2}$ respectively. 
This right temperature $T_R$ in (\ref{TLRN-A}) matches the right temperature in 
 \cite{Chen-RN} after considering the unit length for the uplifted extra dimension. 
\par 
The second copy of deformed hidden conformal symmetry generators for the \rn black hole can be obtained from the branch b generators for the Kerr-Newman black hole in appropriate limit of $a=0$.  We notice that $\sigma = -2\pi T_L$ and so the
deformed hidden conformal generators for the \rn read as,
\begin{eqnarray}
 L_ \pm ^b  &=& e^{ \pm \left( {\frac{{\left( {1 + \kappa } \right)\pi QT_R }}{{\left( {1 - \kappa } \right)M}}} \right)t \mp \left( {2\pi T_L } \right)\chi } \left( { \mp \sqrt \Delta  \partial _r  + \left( {\frac{M}{{2\pi r_ +  }}\left( {\kappa r_ +   - r_ -   + r\left( {1 - \kappa } \right)} \right) - \frac{{Q^2 }}{{4\pi r_ + ^2 }}\left( {r_ +   - r_ -  } \right)} \right)\partial _t } \right. \nonumber\\ 
 &&\left. { + \frac{{Mr_ +  }}{{Q^2 2\pi T_R }}\left( {\kappa r_ +   - r_ -   - r\left( {\kappa  - 1} \right)} \right)\partial _\chi  } \right) ,\label{lpmrnb}\\ 
 L_0^b  &=& \frac{{M^2 r_ +  \left( {\kappa  - 1} \right)}}{{\pi Q^3 T_R }}\partial _t  + \frac{{Mr_ +  \left( {\kappa  - 1} \right)}}{{2\pi Q^2 T_R }}\partial _\chi   ~.\label{l0rnb}
 \end{eqnarray}
%\ee
\par 
We also note that for the special value of deformation parameter $\kappa = r_-/r_+$, the generators (\ref{lpmrna}),(\ref{l0rna}),(\ref{lpmrnb}) and (\ref{l0rnb}) reduce to the hidden conformal generators for the \rn black hole \cite{Chen-RN} after setting the unit length for the uplifted extra dimension. 

However, we notice that the hidden conformal generators  (\ref{lpmrna}),(\ref{l0rna}),(\ref{lpmrnb}) and (\ref{l0rnb}) with $\kappa=r_+/r_-$ for the  Reissner-Nordstrom black holes do not simply reduce to the conformal generators for Schwarzschild black holes  \cite{Bertini} by setting $Q=0$. In this limit, as it is clear from table 2, the coefficients $\rho$ and $\sigma$ do not have any finite values, though the other four coefficients are well-defined. The situation is similar in reduction of hidden conformal generators of Kerr-Sen black holes to hidden conformal generators of Gibbons-Maeda-Garfinkle-Horowitz-Strominger black holes \cite{Ghezelbash} or reduction of conformal generators for Kerr black holes \cite{Castro} to Schwarzschild black holes by setting $a=0$. 
To overcome this problem, as it was noticed in \cite{Ghezelbash,Lowe}, we set $\sigma=0$ for the neutral black holes and so the equations (\ref{rhosig1}) and (\ref{rhosig2}) become 
\be 
\rho C_1 + M = 0~~,~~1+\rho\gamma =0.
\ee  
We note that $\rho$, $C_1$, and $\gamma$ contain the free deformation parameter $\kappa$. In branch b, we choose 
$\kappa=-1$ and our deformed conformal generators reduce exactly to those derived in \cite{Bertini} by the mapping
\be 
L^b_0  =  - iH_0~~,~~ L^b_ \pm   = iH_ \pm  .
\ee

\section{Conclusions}\label{conc}
In this paper we find an extended family of hidden conformal symmetry for the 
Kerr-Newman black holes that are characterized by deformation parameter $\kappa$. The deformation of the inner horizon of the black hole in the radial equation of scalar field is justified by the fact that the back-reaction of the scalar field on the black hole geometry makes the inner horizon replaced by a null curvature spacelike singulatity.  The deformed hidden conformal generators constructed explicitly in different conformal pictures for the Kerr-Newman black holes. The deformed hidden  conformal symmetry in th $J$ picture reduces to the hidden conformal symmetry of the Kerr-Newman where the deformation parameter $\kappa$ is set $r_-/r_+$.
Moreover, the deformed hidden symmetry generators for the Kerr-Newman provide such generators for the \rn black holes. As it is expected, restoring the deformation parameter $\kappa$ to be $r_-/r_+$ gives the hidden conformal symmetry generators for the \rn .  
We also support the deformed Kerr-Newman/CFT correspondence by finding the absorption cross section of charged scalars in the Kerr-Newman background. We find perfect agreement  between the gravitational absorption cross section and 2D CFT cross section in three different conformal pictures for the Kerr-Newman black holes.

\vspace*{10mm}

\noindent {\large{\bf Acknowledgments}}

This work was supported by the Natural Sciences and Engineering Research Council of Canada. \newline

\vspace*{5mm}

%%%%% references %%%%%

\end{document}